\documentstyle[prd,aps]{revtex}
\newcommand{\beq}{\begin{equation}}
\newcommand{\beqn}{\begin{eqnarray}}
\newcommand{\eeq}{\end{equation}}
\newcommand{\eeqn}{\end{eqnarray}}
\begin{document}
\draft
\preprint{HUTP-97/A005; hep-ph/9702244}
\title{Preheating in an Expanding Universe:\\
Analytic Results for the Massless Case}
\author{David I. Kaiser}
\address{Lyman Laboratory of Physics, Harvard University, Cambridge, MA
02138 USA}
\date{2 February 1997; Revised 20 April 1997}
\maketitle
\begin{abstract}
Analytic results are presented for preheating in both
flat and open models of chaotic inflation, for the case of massless
inflaton decay into further inflaton quanta.  It is demonstrated that
preheating in both these cases closely resembles that in Minkowski
spacetime.  Furthermore, quantitative differences between preheating in
spatially-flat and open models of inflation remain of order $10^{-2}$ for
the chaotic inflation initial conditions considered here.\footnote{e-mail
address:  dkaiser@fas.harvard.edu}
\end{abstract}
\pacs{PACS 98.80Cq, 04.62.+v; \hspace*{1cm}  HUTP-97/A005;
hep-ph/9702244. Forthcoming in {\it Phys. Rev. D}}
\vskip2pc

\section{Introduction} 
\indent  Recently, a new view of the post-inflationary reheating period
has been established. \cite{KLS1,STB}  In place of the original view, in
which the inflaton decayed perturbatively \cite{oldreh}, an inherently
nonperturbative, 
highly-efficient resonance has been investigated.  The new theory of 
reheating now involves three distinct stages:  an oscillating inflaton
sets up a parametric resonance in its decay to some boson species; this
explosive stage 
has been termed \lq\lq preheating."~\cite{KLS1} (See also 
\cite{STB,DB1,Yosh,DKprd,KT2,Son,DB205,DBfrw,AllCamp,Kim,Zim,kofman}.) 
Next, 
these far-from-equilibrium decay products themselves interact and decay,
as can be studied along the methods in~\cite{oldreh}.  And finally, the
decay products thermalize, completing the reheating process.  Such
explosive preheating can radically change the thermal history of the
universe following inflation; some non-standard effects associated with
the new preheating picture include the possibility for non-thermal
symmetry restoration \cite{KLS2}, supersymmetry breaking and
GUT-scale baryogenesis \cite{KLR}, and
the amplification of gravitational radiation following the preheating
period~\cite{KT1}. \\
\indent  Because preheating is a non-linear, non-equilibrium process,
particle 
production in most specific models must be analyzed numerically, and 
several results have been reported for both Minkowski and spatially-flat
expanding 
spacetimes. \cite{DB1,KT2,DB205,DBfrw,ProRo}  Much of the analytical 
literature to 
date has treated the expanding background case by means of two 
approximations:  (1) that the oscillating zero-mode of the inflaton
oscillates as 
a cosine function, $\varphi_o \propto \cos (mt)$, with $m$ the mass of
the 
inflaton, and (2) that the expansion may be neglected for frequencies $m
\gg H$, where $H$ is the Hubble parameter.  When these approximations are 
made, the equation of motion for the quantum fluctuations (associated
with 
particle production) can be cast in the form of a Mathieu equation.
Solutions 
of the Mathieu equation generically exhibit an infinite hierarchy of
resonance 
bands; for wavenumbers $k$ within these resonance bands, the solutions 
grow exponentially in time, driving the explosive, resonant particle 
production.  As discussed in~\cite{DB205}, however,
in the context of Minkowski 
spacetime, the first of these approximations can lead to large 
quantitative errors when a quartic self-coupling exists for the inflaton.
In these cases, the zero-mode evolves as an elliptic function in time,
and solutions for the fluctuating fields reveal only one single resonance
band in $k$. \\
\indent  In this paper, we extend this analytical study to the case of
inflaton 
decay into inflaton bosons in an expanding Friedmann-Robertson-Walker 
spacetime.  For a quartic interaction potential and a {\it massless}
inflaton, 
the time-evolution of the entire system (zero-mode, quantum fluctuations, 
and background spacetime) can be studied consistently with analytical 
methods.  Because in this case the observed spectrum of cosmic microwave 
background anisotropies places strict limits on the self-coupling,
$\lambda 
\sim 10^{-12}$, we may study this nonperturbative, non-equilibrium
system 
by means of analytic approximations to the full, non-linear equations of 
motion.  In~\cite{DB205}, the large-$N$ approximation is employed to
study 
preheating analytically in Minkowski spacetime.  Here, we rely on a
Hartree 
factorization to study preheating in an expanding spacetime.  As 
demonstrated explicitly in~\cite{DB396}, these two approximation schemes 
are closely related, and in the case of preheating amount only to the 
substitution $\lambda \rightarrow 3 \lambda$ in the equation of motion
for 
the fluctuating field.  (See also~\cite{Coop}.)  Numerical results have 
indicated that the Hartree approximation can break down for quantitative 
results far from the weak-coupling range (that is, for inflaton decay
into a 
distinct species of boson, with coupling constant $g \gg 
\lambda$)~\cite{KT2}.  However, for the case studied here, of inflaton
decay 
into other inflaton quanta, the Hartree approximation should remain 
quantitatively reliable.  With this analytical framework, it is also easy
to 
understand numerical results \cite{KT2,ProRo} which indicate that {\it
no} 
parametric resonance can occur for a massive inflaton in expanding 
spacetime, if its only decay channel is into inflaton quanta. \\
\indent  We present analytic results for preheating for
both 
ordinary chaotic inflation and for chaotic inflation within a model of
open 
inflation. \cite{open}  The growth of the backreaction from the created
quanta 
on the oscillating zero-mode is calculated, as well as the maximum number 
of quanta produced during preheating.  As discussed below, for preheating
of 
a massless inflaton into massless inflaton quanta, the Ricci curvature
scalar 
vanishes during preheating.  Thus, the preheating dynamics for this model
in 
an expanding spacetime remain conformally equivalent to the Minkowski 
spacetime case; it is not surprising, therefore, that the results
presented here 
share the same form as many of the Minkowski results studied 
in~\cite{DB205}.  The change from large-$N$ to Hartree methods, however, 
does change some of the details of the quantitative analysis.  Also, as 
demonstrated below, the spectrum of produced quanta in the open inflation 
scenario takes the same form as that for quanta produced in an expanding, 
spatially-flat model; for chaotic inflation initial conditions, there are
only 
very small numerical deviations between the two situations.  \\
\indent In section II, we present the specific model to be analyzed here,
discuss the approximation scheme to track the quantum backreaction, and 
choose an appropriate initial
vacuum state with reference to which we can measure the resonant particle
production.  Section III includes analysis of the background spacetime
dynamics during the preheating phase.  In section IV, we solve
consistently for both the oscillating zero-mode of the inflaton field and
the coupled quantum fluctuations, for both the flat and open universe
cases.  In section V, we compare numerically the preheating spectra for
flat 
and open inflation, and provide further consistency checks on some of the 
approximations made throughout the analysis.  Concluding remarks follow 
in section VI.  \\

\section{Dynamics of the Model} 
\indent We only consider the case of
inflaton decay into inflaton bosons, due to a non-linear self-coupling.
The Lagrangian density thus may be written:  
\beqn \nonumber {\cal L} &=& \sqrt{-g} \left[ \frac{1}{16\pi G} R -
\frac{1}{2} g^{\mu\nu} \partial_{\mu} \phi \partial_{\nu} \phi - V (\phi
) \right] , \\ 
V(\phi ) &=& \frac{1}{2}  m^2 \phi^2 + \frac{\lambda}{4} \phi^4 .  
\label{L} 
\eeqn 
We will consider the case of an additional non-minimal 
coupling between the inflaton and $R(t)$, the Ricci
curvature scalar, in section III.  The line element for a
general Friedmann-Robertson-Walker (FRW) spacetime may be written
\beq 
ds^2 = -dt^2 + a^2 (t) h_{ij} dx^i dx^j ,
\label{ds2} 
\eeq 
with 
\beqn 
\nonumber h_{ij}
dx^i dx^j &=& d {\bf x}^2 \>\>\> (K = 0) , \\ 
&=& dr^2 + \sinh^2 r \left(
d\theta ^2 + \sin^2 \theta d \phi^2 \right)  \>\>\> (K = -1) ,
\label{h_ij} 
\eeqn 
and $K = 0$ for a flat universe, $K = -1$ for an open
universe.  (Naturally, the angular coordinate $\phi$ should not be
confused with the inflaton field $\phi (t,\> {\bf x} )$.)  In this
metric, the equations of motion follow:  
\beqn
\nonumber H^2 &+& \frac{K}{a^2} = \frac{8\pi G}{3} \left[ \frac{1}{2}
\dot{\phi}^2 + V (\phi) \right] \\ 
\ddot{\phi} &+& 3H \dot{\phi} -
\frac{1}{a^2} \nabla^2 \phi + \frac{d V}{d\phi} = 0 , 
\label{eom1} 
\eeqn
where $H \equiv \dot{a}/a$ is the Hubble parameter, $\nabla^2$ is the
comoving Laplacian operator, and dots represent derivatives with respect
to cosmic time, $t$. \\ 
\indent Making the familiar decomposition between
the \lq\lq classical"  and fluctuating portions of the inflaton field,
\beq 
\phi (t, {\bf x}) = \varphi (t) + \delta
\phi (t, {\bf x}) , 
\eeq 
we see that the potential of equation (\ref{L}) 
contains a coupling between $\varphi$ and $\delta \phi$.  It is this
coupling which, under certain conditions, will allow for a highly
efficient transfer of energy from the \lq\lq classical" to the
fluctuating
portions of the inflaton field; this transfer of energy is manifested as
a
rapid production of out-of-equilibrium inflaton quanta.  Because the
self-coupling strength $\lambda$ is constrained to be very weak for this
model ($\lambda \sim 10^{-12}$), based on the observed anisotropies in
the cosmic microwave background radiation, we may employ the Hartree
factorization to study the non-equilibrium dynamics of these coupled
systems.~\cite{Hartree} This factorization amounts to making a particular
choice of vacuum state (as discussed below), and making the following
substitutions:  
\beqn 
\nonumber (\delta \phi)^3
&\rightarrow& 3\langle (\delta \phi )^2 \rangle (\delta \phi ), \\ 
(\delta
\phi )^4 &\rightarrow& 6 \langle (\delta \phi)^2 \rangle (\delta \phi)^2
- 3 \langle (\delta \phi^2) \rangle^2 , 
\label{hartree} 
\eeqn 
where
quantities in brackets indicate expectation values of the associated
quantum operators (to be defined explicitly below); the tadpole condition
further requires $\langle (\delta \phi ) \rangle = 0$.  \\ 
\indent In an
open universe, there exists a physical curvature length-scale
$a(\eta)/\vert K \vert$, with $K = -1$.  The corresponding comoving
curvature scale is thus simply $+1$.  As usual, scalar fields may be
expanded in eigenfunctions of the comoving Laplacian.  Writing $\nabla^2
f_P ({\bf x})  = - P^2 f_P ({\bf x})$, then in an open universe, modes
with eigenvalue $P^2 \geq 1$ vary on scales shorter than the comoving
curvature scale, and hence may be labelled \lq\lq subcurvature" modes,
whereas modes with $0 \leq P^2 < 1$ correspond to \lq\lq supercurvature" 
modes. \cite{LythWoc,GarBel} As first noted in~\cite{DKsup}, and as
demonstrated below in section V, preheating in an open universe will {\it
only} populate subcurvature modes for potentials of the form in equation
(\ref{L}).  For this reason, we may expand the fluctuating field $\delta
\phi$ as follows:  
\beq 
\delta \phi (t,\> {\bf x}) = \int d\tilde{\mu} (k) \> \delta \phi_{kjm}
(t) \> Z_{kjm} (\bf{x}) ,
\label{chi_k} 
\eeq 
where the eigenfunctions of the Laplacian obey
\beq 
\nabla^2 Z_{kjm} ({\bf x}) = - (k^2 - K) 
Z_{kjm} ({\bf x}) .  
\label{Zkjm} 
\eeq 
Here $k$ is the comoving {\it subcurvature} wavenumber, with $0 \leq k^2
< \infty$; it is related to the
eigenvalue $P^2$ above by $k^2 = P^2 + K$, and only applies for $P^2 \geq
1$.  The measure for the wavenumber integral is
\cite{LythWoc,GarBel,BirDavies} 
\beqn
\nonumber \int d\tilde{\mu} (k) &=& \int_0^{\infty} dk \> k^2 \>\>\>
\>\>\>\> (K = 0) ,\\ 
&=& \int_0^{\infty} dk \sum_{j = 0}^{\infty} \sum_{m = -j}^{j} 
\>\>\>\>\>
(K = -1).
\label{dmu(k)}
\eeqn
With these definitions, the eigenfunctions of the Laplacian may be
written:
\beqn
\nonumber Z_{kjm} ({\bf x}) &=& (2\pi)^{-3/2} e^{i {\bf k} \cdot 
{\bf x}} \>\>\> \>\>\>\> (K = 0) ,\\
\nonumber  Z_{kjm} (r,\> \theta ,\> \phi ) &=& \sqrt{\frac{2}{\pi}} 
\prod_{s = 0}^{j} \left( s^2 + k^2 \right)^{-1/2} Y_{jm} (\theta ,\>
\phi) \\
 &\times& (\sinh r )^j \left( \frac{ -1}{\sinh r} \frac{d}{dr}
\right)^{j + 1} \cos (kr) \>\>\> (K = -1) ,
\label{Zkjm2}
\eeqn
with $Y_{jm} (\theta , \> \phi )$ the usual spherical harmonics. 
\cite{GarBel,BirDavies,eigen}  \\
\indent To study particle production during the preheating epoch, we may
now promote the field $\delta \phi$ to a Heisenberg operator: 
\beq 
\hat{\delta \phi} (t , \>{\bf x}) = \int
d\tilde{\mu} (k) \left[ \delta\phi_k (t) \>\hat{a}_{kjm}\> Z_{kjm} ({\bf
x}) + \delta\phi^*_k (t) \>\hat{a}_{kjm}^{\dagger} \> Z^*_{kjm} ({\bf x})
\right] ,
\label{hat-delta-phi} 
\eeq 
with $\hat{a}_{kjm}$ and
$\hat{a}_{kjm}^{\dagger}$ the canonical time-independent annihilation and
creation operators, respectively.  These are defined with respect to the
initial Fock vacuum.  With this, the expectation value $\langle
\hat{\delta \phi}^2 \rangle$ becomes, for both $K = 0$ and $K = -1$,
\beq 
\langle \hat{\delta \phi}^2 
\rangle = \int_{0}^{\infty} \frac{dk \> k^2}{2\pi^2} \left| \delta
\phi_k (t) \right|^2 , 
\label{delta-phi-2} 
\eeq 
where translational
invariance allows $\langle \hat{\delta\phi}^2 \rangle$ to depend only on
time; we will therefore write this quantity as $\langle \hat{\delta
\phi}^2 (t) \rangle$.  Defining the constant 
\beq 
{\cal C}^2 \equiv \langle
\hat{\delta\phi}^2 (t_o) \rangle , 
\label{C2} 
\eeq 
where $t_o$ is the
beginning of the preheating epoch,
and switching to conformal time, $d\eta \equiv a^{-1} dt$, we may
introduce the following dimensionless variables:  
\beqn 
\nonumber \tau &=& {\cal C} \eta \>\>,\>\> \ell = k/{\cal C} \>\> , \>\>
M = m / {\cal C} \>, \\
\nonumber \varphi (t) &=& \frac{{\cal C}}{\sqrt{\lambda}} \frac{\psi
(\tau)}{a (\tau)} \>\>,\>\> \delta \phi_k (t) = \frac{1}{\sqrt{{\cal
C}}} \frac{\chi_{\ell} (\tau)}{a (\tau)} , \\ 
\Sigma (\tau) &=& \left[\langle \chi^2 (\tau) \rangle - \langle \chi^2 
(\tau_o) \rangle \right] . 
\label{dless} 
\eeqn 
Then $\langle \hat{\delta\phi}^2 (t) \rangle = {\cal
C}^2 a^{-2} (\tau)  \langle \chi^2 (\tau) \rangle$, and using the
factorization of equation (\ref{hartree}), the equations of motion for
the $\psi$ and $\chi$ fields may be written:  
\beqn 
\nonumber \left[
\frac{d^2}{d \tau^2} + a^2 (\tau) M^2  - \frac{a^{\prime\prime}}{a} +
\psi^2 (\tau) + 3 \lambda a^2
(\tau_o) + 3 \lambda \Sigma (\tau) \right] \psi (\tau) &=& 0 , \\ 
\left[
\frac{d^2}{d \tau^2} + \ell^2 - {\cal K} + a^2 (\tau) M^2  -
\frac{a^{\prime\prime}}{a} + 3 \psi^2 (\tau) + 3 \lambda a^2 (\tau_o) + 3
\lambda \Sigma (\tau) \right] \chi_{\ell} (\tau) &=& 0 , 
\label{eomdless}
\eeqn 
where primes denote derivatives with respect to $\tau$, and ${\cal
K} \equiv {\cal C}^{-2} K$.  It is convenient to define the frequencies
$W_{\ell} (\tau)$ and $\omega_{\ell} (\tau)$ as 
\beqn 
\nonumber W_{\ell}^2 (\tau) &\equiv&
\ell^2 - {\cal K} + a^2
(\tau) M^2 - \frac{a^{\prime\prime}}{a} , \\ 
\omega_{\ell}^2 (\tau)
&\equiv& W_{\ell}^2 (\tau) + 3 \psi^2 (\tau) + 3 \lambda a^2 (\tau_o) + 3
\lambda \Sigma (\tau) .  
\label{Well} 
\eeqn 
The quantity
$\Sigma (\tau)$ measures the backreaction of created quanta on the
evolution of the oscillating zero-mode, $\psi (\tau)$, and $\Sigma
(\tau_o) = 0$.  We will study solutions of the coupled equations of
equation (\ref{eomdless}) in section IV, for early times when $\lambda
\Sigma (\tau)$ may be neglected.  We will also determine
self-consistently in that section the time at which the
approximation $\lambda \Sigma (\tau) \rightarrow 0$ breaks down. \\ 
\indent It remains in this section to
derive an expression for the particle number operator appropriate for the
non-equilibrium dynamics.  In terms of the rescaled field $\chi$, the
quantum fluctuation operator $\hat{\delta\phi}$ may be written: 
\beq 
\frac{1}{\sqrt{ {\cal C}}}\> \hat{\delta\phi}
(t,\> {\bf x}) = \int d\tilde{\mu} (\ell)  \frac{1}{a (\tau)} \left[
\chi_{\ell} (\tau) \>\hat{a}_{\ell jm}\> Z_{\ell jm} ({\bf x}) +
\chi_{\ell}^* (\tau) \> \hat{a}_{\ell jm}^{\dagger} \> Z_{\ell jm}^* ({\bf
x}) \right] ,
\label{delta-phi-hat} 
\eeq 
and, from the Lagrangian density in equation
(\ref{L}), the Heisenberg operator for the conjugate field may be
written: 
\beq 
\sqrt{ {\cal C}} \> \hat{\Pi} (t,
\> {\bf x}) = \int d\tilde{\mu} (\ell) a (\tau) \sqrt{h ({\bf x})} \left[
\left(\chi_{\ell}^{\prime} (\tau) - {\cal H} (\tau) \chi_{\ell} (\tau)
\right) \> \hat{a}_{\ell jm} \> Z_{\ell jm} ({\bf x}) + {\rm h.c.}
\right] ,
\label{Pi} 
\eeq 
with ${\cal H} (\tau) \equiv a^{\prime}/a$, and \lq\lq
h.c." denoting hermitian conjugate.  Canonical quantization then gives
the normalization condition for the mode functions $\chi_{\ell} (\tau)$: 
\beq 
\chi_{\ell} \> \chi_{\ell}^{*\prime} -
\chi_{\ell}^{\prime} \> \chi_{\ell}^* = i .
\label{norm}
\eeq
This normalization comes from quantizing the fields $\hat{\delta \phi}$
and $\hat{\Pi}$ on a Cauchy surface.  Yet for models of open inflation,
the interior of the nucleated bubble is {\it not} a Cauchy surface for
the
entire de Sitter space.  It has been demonstrated, however, that for
scalar fields expanded only in subcurvature modes, quantizing {\it as if}
the interior of the nucleated bubble were a proper Cauchy surface
reproduces exactly the same result as when the fields are quantized on a
proper Cauchy surface and analytically continued inside the
bubble.~\cite{cauchy} For this reason, we may use the normalization in
equation (\ref{norm}) for both the flat and open inflation cases. In the
context of preheating, this makes sense physically, since we are
quantizing these fields after the volume of the interior of the bubble
has
expanded by a factor of $\sim \left( e^{65} \right)^3$ (according to any
\lq\lq observers" on the interior of the bubble), so that all causal
properties of these fields should be specifiable with reference to the
interior of the bubble alone. \\ 
\indent To calculate the particle number per mode, it is convenient to
write 
\beqn
\nonumber \hat{\delta \phi} (t,\>{\bf x}) &=& \int d\tilde{\mu} (\ell) 
\hat{\delta\phi}_{\ell jm} (t) Z_{\ell jm} ({\bf x}) , \\ 
\hat{\Pi} (t,\>
{\bf x}) &=& \int d\tilde{\mu} (\ell) \hat{\Pi}_{\ell jm} (t,\> {\bf x})
Z_{\ell jm} ({\bf x}) .  
\label{hat-delta-phi2} 
\eeqn 
For $K = 0$, we may thus write
\beqn 
\nonumber \frac{1}{\sqrt{ {\cal C}}} \>
\hat{\delta \phi}_{\ell jm} (t) &=& \frac{1}{a (\tau)} \left[ \chi_{\ell}
\> \hat{a}_{\vec{\ell}} + \chi_{\ell}^* \>
\hat{a}_{-\vec{\ell}}^{\dagger} \right] , \\ 
\sqrt{{\cal C}}\> \hat{\Pi}_{\ell jm} (t,\>{\bf x}) &=& a
(\tau)  \sqrt{h({\bf x})} \left[ \left( \chi_{\ell}^{\prime} - {\cal H}
\chi_{\ell} \right)  \hat{a}_{\vec{\ell}} + \left(\chi_{\ell}^{*\prime} -
{\cal H} \chi_{\ell}^* \right) \hat{a}_{ - \vec{\ell}}^{\dagger} \right] . 
\label{flatfourier} 
\eeqn 
As pointed out in~\cite{GarBel} for the $K = -1$
case, we may use the purely-real representation of the $Y_{jm} (\theta ,
\> \phi )$'s, which makes the eigenfunctions $Z_{\ell jm} ({\bf x})$
purely real for subcurvature modes.  Thus, for $K = -1$ we also may write
expressions as in equation (\ref{flatfourier}), but with the arguments of
the creation and annihilation operators changed to:
$\hat{a}_{\vec{\ell}}
\rightarrow \hat{a}_{\ell jm}$ and $\hat{a}_{-\vec{\ell}}^{\dagger}
\rightarrow \hat{a}_{\ell jm}^{\dagger}$. \\ 
\indent Care must be taken
when studying preheating in an expanding universe to choose an
appropriate
vacuum state with respect to which we can measure particle production. 
There are two independent concerns:  first, preheating is a
non-equilibrium process among interacting fields, so defining \lq\lq
particle" states is ambiguous even in Minkowski spacetime~\cite{DB205}; 
in addition, we have the usual ambiguity in defining \lq\lq free
particle"
states in any non-Minkowskian spacetime~\cite{BirDavies}.  At $\tau
\rightarrow - \infty$ (where we take the bubble nucleation to have
occurred
at $\tau = - \infty$), the spacetime inside the nucleated bubble is a de
Sitter spacetime, so in the absence of interactions, the vacuum state for
the $\delta \phi$ field should be the Bunch-Davies vacuum.  If the
transition from pure de Sitter expansion to the different rate of
expansion at the time of preheating occurs adiabatically, then at the
onset of preheating the $\chi_{\ell}$ modes should obey the initial
conditions:  $\chi_{\ell} (\tau_o) = (2 W_{\ell} (\tau_o))^{-1/2}$ and
$\chi_{\ell}^{\prime} (\tau_o) = -i (W_{\ell} (\tau_o) / 2)^{1/2}$ in the 
absence
of interactions.  These modes would represent the free-particle \lq\lq
adiabatic" states.~\cite{BirDavies}  Furthermore, if we consider the
interaction strength to be turned on adiabatically beginning some time
before $\tau_o$, then these initial conditions should be replaced by:
\beq
\chi_{\ell} (\tau_o) = \frac{1}{ \sqrt{2 \omega_{\ell} (\tau_o)}} \>\>,
\>\> \left( \frac{d \chi_{\ell}}{d \tau} \right)_{\vert \> \tau = 
\tau_o} = -i \sqrt{ \frac{\omega_{\ell} (\tau_o)}{2}} .
\label{initial}
\eeq
Equation (\ref{initial}) gives the initial conditions for the \lq\lq
adiabatic" particle states for the non-equilibrium dynamics at the onset
of preheating. We may therefore define adiabatic creation and
annihilation operators $\hat{\alpha}_{\ell jm} (\tau)$ and
$\hat{\alpha}_{\ell jm}^{\dagger} (\tau)$ through the relations: 
\beqn 
\nonumber \frac{1}{\sqrt{{\cal C}}} \> \hat{\delta
\phi}_{\ell jm} (\tau ) &\equiv& \frac{1}{a(\tau)}
\frac{1}{\sqrt{2 \omega_{\ell} (\tau)}} \left[ \hat{\alpha}_{\ell jm}
(\tau) 
e^{-i \int \omega_{\ell}(\tau) d\tau} + \hat{\alpha}_{\ell jm}^{\dagger}
(\tau) e^{i \int \omega_{\ell} (\tau) d\tau } \right] , \\ 
\sqrt{{\cal C}}\> \hat{\Pi}_{\ell jm} (\tau , \> {\bf x}) &\equiv& -i
\sqrt{\frac{\omega_{\ell} (\tau)}{2}} \> a(\tau) \sqrt{h ({\bf x})} 
\left[ \left( 1 - i \frac{ {\cal H} (\tau)}{\omega_{\ell} (\tau)}
\right) \hat{\alpha}_{\ell jm} (\tau) e^{-i \int \omega_{\ell} (\tau)
d\tau} + {\rm h.c.} \right] .
\label{alpha-fourier}
\eeqn
The $\hat{\alpha}_{\ell jm} (\tau)$ annihilates the time-dependent 
adiabatic
vacuum state:  $\hat{\alpha}_{\ell jm} (\tau) \vert 0 (\tau)> = 0$ for all
$\ell$, $j$, and $m$.  These adiabatic creation and annihilation operators
can be related to the time-independent operators $\hat{a}_{\ell jm}$ and
$\hat{a}_{\ell jm}^{\dagger}$ by means of a Bogolyubov transformation.
As above, we replace the arguments of these operators
as $\hat{\alpha}_{\ell jm} \rightarrow \hat{\alpha}_{\vec{\ell}}$ and
$\hat{\alpha}_{\ell jm}^{\dagger} \rightarrow \hat{\alpha}_{-
\vec{\ell}}^{\dagger}$ in equation (\ref{alpha-fourier}) for the $K = 0$
case.  Note that these
expansions are only valid when the frequency $\omega_{\ell} (\tau)$ is
purely real; as demonstrated in section III, this will always be the case
for the scenarios considered here. \\
\indent From the expansions in equations (\ref{flatfourier}) and
(\ref{alpha-fourier}), we may solve for the adiabatic particle number per
mode.  The result is the same for both the $K = 0$ and $K = -1$ cases: 
\beq 
N^{\rm ad}_{\ell} (\tau) = \langle 
\hat{\alpha}_{\ell jm}^{\dagger} (\tau)  \hat{\alpha}_{\ell jm} (\tau)
\rangle = \frac{\omega_{\ell} (\tau)}{2} \left[ \left| \chi_{\ell} (\tau) 
\right|^2 + \frac{1}{\omega_{\ell}^2 (\tau)} \left| \chi_{\ell}^{\prime}
(\tau) \right|^2 \right] - \frac{1}{2} .
\label{Nad} 
\eeq 
This yields the number of \lq\lq adiabatic"-state 
quanta produced relative to the initial Fock
vacuum, $\vert 0 > = \vert 0 (\tau_o) >$.
With this choice
of vacuum state and initial conditions, the particle number for inflaton
quanta does indeed vanish at the onset of preheating.  As discussed in
section III, for the models of interest here, the background spacetime
will evolve as if it were radiation-dominated for the entire period of
preheating, so that there will be no further Bogolyubov transformations
needed to relate the particle number at the beginning to that at the end
of preheating.  \\ 
\indent With the choice of initial conditions for the
$\chi$ field, equation (\ref{initial}), the quantity ${\cal C}^2 =
\langle
\hat{\delta\phi}^2 (\tau_o) \rangle$ is formally quadratically divergent,
and would have to be renormalized with some regularization scheme.  (This
is essentially the zero-point energy divergence.)  For here, we will
simply note that to remain consistent, we require the energy density of
the \lq\lq classical" portion of the inflaton field to exceed that of the
quantum fluctuation portion at the beginning of preheating; that is,
$\varphi^2 (\tau_o) \gg \langle \hat{\delta \phi}^2 (\tau_o) \rangle$. 
This is equivalent to the requirement
\beq 
\psi^2 (\tau_o) \gg \lambda a^2 (\tau_o) ,  
\label{psi_o} 
\eeq 
using the definition of $\psi (\tau)$ in equation (\ref{dless}).  We will
make use of this relation below. \\

\section{Evolution of the Background Spacetime} 
\indent In this section we
study the background spacetime dynamics during the period of preheating. 
Up until the time $\tau_o$, the fluctuations of the field $\delta \phi$
are in their vacuum state, and we need only consider the energy density
from the \lq\lq classical" part of the inflaton field, $\varphi$.  In
terms of cosmic time $t$, this may be written:  
\beq 
\rho_{\varphi} (t) = \frac{1}{2} \dot{\varphi}^2 + \frac{1}{2} m^2
\varphi^2 + \frac{1}{4} \lambda \varphi^4 .  
\label{rho1} 
\eeq 
In terms of dimensionless conformal time $\tau$ and the
rescaled field $\psi$, the energy density is thus 
\beq 
\rho_{\varphi} (\tau) = \frac{ {\cal
C}^4}{2 \lambda a^4
(\tau)} \left[ \frac{1}{2} \psi^4 + \psi^{\prime 2} + a^2 M^2 \psi^2 +
{\cal H} \psi \left( {\cal H} \psi - 2 \psi^{\prime}
\right) \right] .  
\label{rho2} 
\eeq 
At the onset of preheating, the field
$\psi$ begins to oscillate (as treated explicitly in section IV), so that
averaging over a period of its oscillations, both $\psi^2$ and
$\psi^{\prime 2}$ will remain nearly constant.  For the chaotic inflation
scenario we consider here, then, when the mass $m$ vanishes, the energy
density at the onset of preheating will be dominated by the $a^{-4}$
terms.  Note that by working in terms of conformal time, we do not
have to make any assumptions about the magnitude or rate of change of 
the Hubble parameter $H (t)$, as we would if we worked in terms of
cosmic time, $t$.  Instead, we only need to use the chaotic inflation 
initial conditions, $\psi^4 (\tau_o) \gg \psi^2 (\tau_o)$.  We will
confirm 
in section V that when $m = 0$, these initial conditions ensure that
${\cal H} \ll \psi$ during preheating. \\
\indent  Thus, when $m = 0$ (and only then), we may
approximate the time-evolution of the background spacetime as that of a
radiation-dominated FRW metric at the very onset of preheating.  (This
point is also noted in~\cite{kofman}.)  Furthermore, for the case of
inflaton decay into inflaton bosons, the produced quanta will also be
massless when $m = 0$, so that over the entire preheating period we may
keep the approximation $\rho (\tau)  \propto a^{-4} (\tau)$.  Note that
this behavior of the energy density holds even though the quanta are far
from thermal equilibrium when produced. \\ 
\indent Using the Friedmann
equation (see equation (\ref{eom1})), and writing $\rho (\eta) = \rho_o
(a (\eta) / a_o)^{-4g}$, we may solve for the behavior of the scale factor
in terms of conformal time $\eta$ for both the $K = 0$ and $K = -1$ cases. 
For a flat universe, the scale factor evolves as:  
\beq 
a (\eta) = a_o \left( \frac{ \eta}{\eta_o} \right)^{1/(2g -
1)} \>\>,\>\> g \neq \frac{1}{2} , 
\eeq 
and for $K = -1$,
\beq 
a(\eta) = a_o \sinh^d \left(
\frac{\eta}{\vert d \vert} + X \right)  \>\>,\>\> d \equiv \frac{2}{4g -
2}\>\>,\>\>g \neq \frac{1}{2} .  
\eeq 
In this case, by defining $a_o = a(\eta_o)$, the Friedmann
equation further implies the relation $H_o^2 = 2/a_o^2$.  In terms of the
dimensionless conformal time $\tau$, setting $g = 1$ (for $m = 0$) yields
the very simple result for $K = 0,\> 1$:
\beq
\frac{a^{\prime\prime}}{a} = - {\cal K} . 
\label{aprimeprime} 
\eeq
This result, valid only for $m = 0$, means that the addition of a
non-minimal coupling between the inflaton and the Ricci curvature scalar
of the form $\frac{1}{2} \xi R \phi^2$ would not affect any of the
preheating dynamics.  To see this, consider the Ricci curvature scalar in
terms of cosmic time, $t$:
\beq
R (t) = \frac{6}{a^2 (t)} \left[ \ddot{a} a + \dot{a}^2 + K \right] .
\label{R1}
\eeq
Rewriting this in terms of the dimensionless conformal time $\tau$,
\beq
R (\tau) = \frac{6 {\cal C}^2}{a^2 (\tau)} \left[
\frac{a^{\prime\prime}}{a} + {\cal K} \right] ,
\label{R2}
\eeq
we see from equation (\ref{aprimeprime}) that a consistent solution of the
modified Friedmann equation would include $R = 0$ for both $K = 0$ and $K
= -1$ when $m = 0$.  Thus, for the massless case, a non-minimal coupling
would not affect the preheating dynamics. \\
\indent  Using equation (\ref{aprimeprime}), we may further confirm
that the frequency $\omega_{\ell} (\tau)$ will always remain real (see
equation (\ref{Well})), which allows us to employ the adiabatic creation
and annihilation operators of equation (\ref{alpha-fourier}).  In fact,
when $m = 0$, this frequency is the same for both the $K = 0$ and $K = -1$
cases: 
\beq 
\omega_{\ell}^2 (\tau) = \ell^2 +
3\psi^2 (\tau) + 3\lambda a^2 (\tau_o) + 3 \lambda \Sigma (\tau) . 
\label{omega_ell2} 
\eeq 
The condition of equation (\ref{psi_o}) further
allows us to neglect the $3 \lambda a^2 (\tau_o)$ term relative to the $3
\psi^2$ term during preheating.  With these expressions for the evolution
of the background spacetime, we may now study the dynamics of the $\psi$
and $\chi$ fields during preheating. \\

\section{Evolution of the Fields during Preheating} 
\indent Using equations (\ref{psi_o}) and (\ref{aprimeprime}),
we may rewrite the equation of motion for the $\psi$ field in equation
(\ref{eomdless}) as 
\beq 
\left[
\frac{d^2}{d\tau^2} + \psi^2 (\tau) + {\cal K} + 3 \lambda \Sigma (\tau)
\right] \psi (\tau) = 0 .  
\label{psieom} 
\eeq 
In this section, we will
shift the time of the onset of preheating to $\tau_o = 0$.  With the
initial conditions, $\psi (0) = \psi_o$ and $(d\psi /d \tau )_{\vert \tau
= 0} = 0$, the
equation of motion for $\psi (\tau)$ may be solved in terms of a Jacobian 
elliptic cosine function for early times, when the backreaction $3\lambda 
\Sigma (\tau)$ is negligible. In general, the Jacobian cosine function
${\rm cn} (u, \> \overline{k})$ obeys the differential
equation~\cite{Bateman}:
\beq 
\left[ \frac{d^2}{du^2} +  \left( 1 - 2 \overline{k}^2 + 2 \overline{k}^2
{\rm c}^2 \right) \right] {\rm c} = 0 , 
\eeq 
where we have used the abbreviation ${\rm c} \equiv {\rm cn}
(u, \> \overline{k})$.  For early times, then, the $\psi$ field
oscillates as 
\beq 
\psi (\tau) = \psi_o \> {\rm cn} \left(
\sqrt{\psi_o^2 + {\cal K}} \> \tau ,\>\frac{\psi_o}{\sqrt{2(\psi_o^2 +
{\cal K}) }} \right) , 
\label{psisol} 
\eeq 
for $\psi_o^2 \geq - 2 {\cal K}$, appropriate for a chaotic inflation
scenario.  \\
\indent  In terms of the time-like variable $u = \sqrt{\psi_o^2 + {\cal
K}} \> \tau$, the equation of motion for the fluctuations becomes:
\beq
\left[ \gamma^2 \frac{d^2 }{du^2} + {\ell}^2 + 3 \psi^2 (u) \right]
\chi_{\ell} (u) = 0 ,
\label{chieom2}
\eeq
for early times, when we may neglect $\lambda \Sigma (\tau)$.  Here we 
have defined the constant $\gamma^2 \equiv \psi_o^2 + {\cal K}$.  We may 
rewrite the Jacobian cosine function in terms of the doubly-periodic 
Weierstrass function $\wp (z)$ by noting both that ${\rm cn}^2 (u,\> 
\overline{k}) + {\rm sn}^2 (u, \> \overline{k}) = 1$, and that 
\beq
{\rm sn}^2 (u, \>\overline{k} ) = \frac{1}{\overline{k}^2 (e_1 - e_3)}
\left[ 
\wp \left(\frac{u + i K^{\prime} (\overline{k})}{\sqrt{e_1 - e_3}}\right) 
- e_3 \right] ,
\label{wp1}
\eeq
where $K^{\prime} (\overline{k})$ is the complementary complete elliptic 
integral of the first kind, and the $e_i$ are the three constants
associated 
with the Weierstrass function $\wp (z)$.~\cite{Bateman}  These constants 
sum to zero:  $e_1 + e_2 + e_3 = 0$; in terms of them, the modulus 
$\overline{k}$ may be written:  $\overline{k}^2 = (e_2 - e_3 ) / (e_1 -
e_3 )$.  The function $\wp (z)$ is periodic with respect to the two
periods, $2\omega$ and $2\omega^{\prime}$, as follows
\cite{Bateman,AbSteg}:
\beqn
\nonumber  \wp (z + 2M \omega + 2N \omega^{\prime} ) &=& \wp (z) , 
\\
\omega = \frac{K (\overline{k})}{\sqrt{e_1 - e_3}} \>\> , \>\>  
\omega^{\prime} &=& i\frac{K^{\prime} (\overline{k})}{\sqrt{e_1 - e_3}} ,
\label{wp2}
\eeqn
for integer $M$ and $N$.  We may set $(e_1 - e_3) = 1$, which yields
\beq
e_1 = \frac{3 \gamma^2 + {\cal K}}{6 \gamma^2} \>\>,\>\> e_2 = - \frac{ 
{\cal K}}{3 \gamma^2} \>\>,\>\> e_3 = - \frac{ (3 \gamma^2 - {\cal K})}{
6 \gamma^2} .
\label{e_i}
\eeq
With these substitutions, equation (\ref{chieom2}) becomes:
\beqn
\nonumber  \left[ \frac{d^2}{du^2} + p^2 - 6 \wp (u + \omega^{\prime}) 
\right] \chi_{p} (u) = 0 , \\
p^2 (k) \equiv \frac{\ell^2 - 2 {\cal K}}{\gamma^2} .
\label{chieom3}
\eeqn
This is now in the form of a Lam\'{e} equation.  We may solve it
explicitly by introducing two dimensionless constants, $a$ and $b$, by
means of the transcendental relations:
\beqn
\nonumber  3 \wp (a) + 3 \wp (b) &=& - p^2 , \\
\wp^{\prime} (a) &=& - \wp^{\prime} (b) ,
\label{ab}
\eeqn
where primes in this section will denote derivatives with respect to $z
\equiv u / \sqrt{e_1 - e_3} = u$.  Using the differential equation 
\cite{Bateman,AbSteg}
\beqn
\nonumber \wp^{\prime 2} (z) &=& 4 \wp^3 (z) - g_2 \wp (z) - g_3 , \\
g_2 &\equiv& 2 \left( e_1^2 + e_2^2 + e_3^2 \right) \>\>,\>\> g_3 \equiv 
4 e_1 e_2 e_3 ,
\label{wp-prime}
\eeqn
the relations in equation (\ref{ab}) imply:
\beqn
\nonumber \wp (b) &=& - \frac{1}{6} p^2 + \frac{1}{2} \sqrt{1 + \frac{ 
{\cal K}^2 }{3 \gamma^4} - \frac{1}{3} p^4 } , \\
\wp^{\prime 2} (b) &=& \frac{4}{27} p^6 - \frac{1}{3} \left( 1 +
\frac{{\cal 
K}^2}{3 \gamma^4} \right) p^2 - \frac{{\cal K}}{3 \gamma^2} \left(1 - 
\frac{{\cal K}^2}{9 \gamma^4} \right) .
\label{wpb}
\eeqn
\indent  With these relations, solutions of equation (\ref{chieom3}) may
be written:
\beq
U_p (u) = \frac{\sigma (u + \omega^{\prime} + a) \>\sigma (u + 
\omega^{\prime} + b) \>\sigma^2 (\omega^{\prime} )}{\sigma^2 (u + 
\omega^{\prime} ) \>\sigma (\omega^{\prime} + a) \>\sigma 
(\omega^{\prime} + b)} \exp \left[ - u \left( \zeta (a) + \zeta (b)
\right) \right] ,
\label{Up1}
\eeq
where $\sigma (z)$ and $\zeta (z)$ are the quasi-periodic Weierstrass 
functions, defined by the relations \cite{Bateman,AbSteg}:
$\zeta^{\prime} 
(z) \equiv - \wp (z)$ and $\sigma^{\prime} (z) / \sigma (z) \equiv \zeta 
(z)$.  The solution $U_p (u)$ is normalized as $U_p (0) = 1$, and the 
linearly-independent solution is $U_p (-u)$.   Using the initial
conditions of 
equation (\ref{initial}), the fluctuations $\chi_{\ell} (\tau)$ can be
written as a linear combination of $U_p (u)$ and $U_p (-u)$:
\beq
\chi_{\ell} (u) = \frac{1}{2 \sqrt{2 \omega_{\ell} (0)}} \left[ \left( 1
+ i 
\frac{\omega_{\ell} (0)}{ \gamma U_p^{\prime} (0)} \right) U_p (-u) + 
\left( 1 - i \frac{ \omega_{\ell}(0)}{\gamma U_p^{\prime} (0)} \right)
U_p (u) \right] ,
\label{chi-U}
\eeq
where the primes here denote $d / du = \gamma^{-1} d / d \tau$.  From the 
form of the number operator, equation (\ref{Nad}), solutions $U_p 
(\pm u)$ which grow in time will contribute to particle production.  \\
\indent  In general, solutions to second-order differential equations
with 
periodic coefficients will obey Floquet's theorem~\cite{Ince}; that is,
for a 
periodic \lq\lq potential" with period $2 \omega$ (as is the case for
equation (\ref{chieom3})), solutions behave as
\beq
U_p (u + 2 \omega ) = U_p (u) e^{i F (p)} ,
\eeq
where the Floquet index $F (p)$ is independent of time.  The solutions
$U_p 
(-u )$ have the Floquet index $- F(p)$.  The quasi-periodicity of the
$\sigma (z)$ functions~\cite{AbSteg},
\beq
\sigma (z + 2 \omega) = - \sigma (z) \exp \left[ 2 \zeta (\omega) \left(
z + \omega \right) \right] ,
\eeq
yields, for the solutions in equation (\ref{Up1}),
\beq
F(p) = 2i \left[ \omega \left( \zeta (a) + \zeta (b) \right) - \left( a +
b \right) \zeta (\omega ) \right] .
\label{Fp}
\eeq
The solutions $U_p (u)$ will reveal exponential instabilities, then,
whenever 
$F(p)$ has a non-zero imaginary component.  Equations (\ref{Nad}) and 
(\ref{chi-U}) reveal that it is these instabilities which are responsible
for 
the highly efficient, resonant particle production during preheating.  As
is clear 
from the relations of equation (\ref{ab}), together with series
expansions of 
$\wp (z)$, $\wp^{\prime} (z)$, and $\zeta (z)$ (see~\cite{AbSteg}), such 
instabilities will occur only when {\it both} $\wp (b)$ and $\wp^{\prime} 
(b)$ are real.  From equation (\ref{wpb}), we see that for the physical 
wavenumbers $0 \leq k^2 < \infty$, there exists only {\it one} band in
which these exponential instabilities may occur:
\beq
\frac{3}{2} \sqrt{1 + \frac{ {\cal K}^2}{3 \gamma^4} } \leq p^2 \leq
\sqrt{3 + \frac{ {\cal K}^2}{\gamma^4} } ,
\label{p2}
\eeq
or, in terms of $\psi_o$ and the dimensionless wavenumber $\ell$:
\beq
\frac{3}{2} \sqrt{ \psi_o^4 + 2 {\cal K} \psi_o^2 + \frac{4}{3} {\cal
K}^2} + 2 
{\cal K} \leq \ell^2 \leq \sqrt{3 \psi_o^4 + 6 {\cal K} \psi_o^2 + 4 
{\cal K}^2} + 2 {\cal K} .
\label{ell2}
\eeq
The existence of only one single resonance band for physical wavenumbers
in this case matches the results found for Minkowski spacetime
in~\cite{DB205}.  This resonance band is in terms of {\it comoving}
wavenumber $k$; because we have neglected interactions amongst the
final-state bosons, the only way a mode $\chi_{\ell} (\tau)$ will slide out
of the resonance band in this system is when the backreaction $\lambda
\Sigma (\tau)$ grows so large that it damps the oscillations of the
zero-mode, halting the parametric resonance.  We 
will demonstrate in section V that for preheating in an expanding open
FRW universe, this range only includes subcurvature modes.  \\
\indent  In order to evaluate quantitatively values of the solutions $U_p 
(\pm u)$ in this resonance band, it is helpful to rewrite the solutions
in 
terms of the quasi-periodic Jacobian theta functions, which possess very 
rapidly-converging series expansions.  Using the relations~\cite{Bateman}
\beqn
\nonumber  \sigma (z) &=& 2 \omega \frac{\vartheta_1 (v)}
{\vartheta_1^{\prime}(0)} \exp \left[ \frac{ \zeta (\omega) z^2}{2
\omega} \right] , \\
\vartheta_1 \left( v + \frac{\omega^{\prime}}{2 \omega} \right) &=& i
\vartheta_4 (v) \exp \left[ - i \pi \left( v + \frac{\omega^{\prime}}{4
\omega} \right) \right],
\eeqn
the solutions $U_p (u)$ may be written:
\beq
U_p (\pm v) = \frac{ \vartheta_4 (v + \alpha ) \>\vartheta_4 (v + \beta ) 
\>\vartheta^2_4 (0)}{\vartheta_4^2 (v) \>\vartheta_4 (\alpha) 
\>\vartheta_4 (\beta)} \exp \left[ \pm i v F (p) \right] ,
\label{Upv}
\eeq
where we have used $2 \omega v \equiv u$, $2\omega \alpha \equiv a$, 
and $2\omega \beta \equiv b$.  The relation~\cite{Bateman}
\beq
\zeta (z) = \frac{\zeta (\omega)}{\omega} z + \frac{1}{2 \omega}
\frac{\vartheta_1^{\prime} (v)}{\vartheta_1 (v)}
\eeq
further allows us to rewrite the Floquet index $F (p)$ as
\beq
F (p) = i \left[ \frac{ \vartheta_1^{\prime} (\alpha)}{\vartheta_1
(\alpha)} + \frac{\vartheta_1^{\prime} (\beta)}{\vartheta_1 (\beta )}
\right] .
\label{Fp2}
\eeq
With the constraints on $a$ and $b$ in equation
(\ref{ab}), the Floquet index will reach a maximum in the resonance band
for $a \rightarrow b$.  We may evaluate the Floquet index at this maximum
by means of the series expansion for $\vartheta_1 (v)$~\cite{Bateman}:
\beqn
\nonumber \frac{\vartheta_1^{\prime} (v)}{\vartheta_1 (v)} &=&
 \pi {\rm ctn} (\pi v) + 4 \pi \sum_{n=1}^{\infty} 
\frac{q^{2n}}{\left(1 - q^{2n} \right)} \sin (2 n \pi v ) , \\
&=& \pi {\rm ctn} (\pi v) + 4 \pi q^2 \sin (2 \pi v) + O (q^4) ,
\label{vartheta-prime}
\eeqn
where the elliptic nome $q \equiv \exp [ - i K^{\prime} (\overline{k}) /
K (\overline{k} )]$ satisfies $0 \leq q \leq 0.0432$.  Writing $F (p)
\simeq 2i ( \vartheta_1^{\prime} (\beta_{\rm max}) / \vartheta_1
(\beta_{\rm max}))$ near its maximum in the resonance band yields 
\beqn
\nonumber  Re \left( \frac{\vartheta_1^{\prime} (\beta_{\rm 
max})}{\vartheta_1
(\beta_{\rm max} )} \right) &=& 4 \pi q + O (q^3 ) , \\
Re \left( \frac{\partial^2}{\partial v^2} \left(
\frac{\vartheta_1^{\prime}
(v)}{\vartheta_1 (v)} \right)_{\vert 
\beta_{\rm max}} \right) &=& - 16 \pi^3 q + O (q^3 ) ,
\label{vartheta2}
\eeqn
where the maximum of the resonance occurs at $\beta_{\rm max}$. From 
the relations in equations (\ref{ab}) and (\ref{wpb}), $\beta_{\rm max}$
corresponds to the wavenumber 
\beq
p^2_{\rm max} \rightarrow \sqrt{3 + \frac{ {\cal K}^2}{\gamma^4} } .
\label{pmax}
\eeq
In other words, the maximum resonance will occur for quanta with
wavenumbers near the top of the resonance band.  This behavior also 
matches that found for Minkowski spacetime in~\cite{DB205}.  \\
\indent  With these expressions, we may determine the growth of the
backreaction due to created quanta, as well as the number of quanta
produced during the preheating epoch.  From the sign of $F (p)$ at its
maximum, it is clear that the $U_p (-u)$ modes will grow exponentially in
time for modes within the resonance band, while the resonant 
$U_p (u)$ modes will decrease exponentially.  Keeping only the growing
modes, then, we may approximate 
\beq
\left| \chi_{\ell} (\tau ) \right|^2 \simeq \frac{1}{8 \omega_{\ell} (0)}
\left| U_p (-u) \right|^2 \left[1 + \frac{ \omega_{\ell}^2 
(0)}{\gamma^2 U_p^{\prime 2} (0)} \right] 
\label{chip}
\eeq
and
\beqn
\nonumber \Sigma (\tau) &\simeq& \int_{0}^{\infty} \frac{d\ell \> 
\ell^2}{2 \pi^2} \left| \chi_{\ell} (\tau) \right|^2 \\
&\simeq& \int_{0}^{\infty} \frac{d \ell \> \ell^2}{16 \pi^2 \omega_{\ell} 
(0)} \left| A_p (u) \right|^2 \left[ 1 + \frac{\omega_{\ell}^2
(0)}{\gamma^2 
U_p^{\prime 2} (0)} \right] \exp \left[ \frac{u}{\omega} Re \left( F (p) 
\right) \right]
\label{Sigma-res}
\eeqn
inside the resonance band.  Here we have written the combination of
oscillating $\vartheta$-functions as the single function, $A_p (u)$.
We will evaluate this integral by means of a 
saddle-point approximation.  Using equation (\ref{chieom3}), we may 
substitute
\beq
d \ell \> \ell^2 = \gamma^3 dp \> p \sqrt{p^2 - \frac{2 {\cal 
K}}{\gamma^2}} ,
\eeq
and using equation (\ref{vartheta2}) we may set
\beq
Re \left( F (p) \right) \simeq 8 \pi q - 16 \pi^3 q \left( p - p_{\rm
max} \right)^2 + O (q^3) .
\eeq
Furthermore, the integral will reach its maximum value when the
oscillating term $A_p (u) = 1$.  This yields
\beq
\Sigma (\tau) \simeq \left[ \frac{\gamma^3 p \sqrt{p^2 - \frac{2 {\cal 
K}}{\gamma^2}} }{128 \pi^3 \omega_{\ell} (0)} \left( 1 + \frac{ 
\omega^2_{\ell} (0)}{\gamma^2 U_p^{\prime 2} (0)} \right) \right]_{\vert 
p_{\rm max}} \frac{1}{\sqrt{q \gamma \tau / \omega}} \exp \left[ \frac{ 
8 \pi \gamma q}{\omega} \>\tau + O(q^3) \right] .
\label{Sigma-res2}
\eeq
Following~\cite{DB205}, we may then write the backreaction in the form
\beq
\Sigma (\tau) \equiv \frac{\gamma^{3/2}}{N ({\cal K}) \sqrt{\tau}} \exp 
\left[ B ({\cal K}) \gamma \tau \right] .
\label{back1}
\eeq
In this form, we may solve for the time $\tau$ when the backreaction $3 
\lambda \Sigma (\tau)$ grows to be of the same magnitude as the
tree-level 
terms.  In our case, this occurs when $3 \lambda \Sigma (\tau) \simeq 3 
\psi^2 (\tau)$. \\
\indent  For the entire range $- 2 {\cal K} \leq \psi_o^2 \leq \infty$,
the 
average of the square of the zero-mode over a period of its oscillations
gives, to a good approximation~\cite{DB205},
\beq
3 \overline{\psi^2 (\tau)} \simeq \frac{3}{2} \psi_o^2 = \frac{3}{2}
\left( \gamma^2 - {\cal K} \right) .
\eeq
Setting the backreaction equal to this quantity yields 
\beq
\tau_{\rm end} \simeq \frac{1}{B (\tilde{\cal K})
 \gamma} \ln \left(\frac{ N (\tilde{\cal K}) \left( 1 - 
3\tilde{\cal K} \right)}{2 \lambda \sqrt{B (\tilde{\cal K}) }}\right) ,
\label{tend}
\eeq
where we have defined $\tilde{\cal K} \equiv {\cal K} / (3 \gamma^2)$.  
Once the backreaction grows to equal the magnitude of the tree-level
terms, 
the parametric amplification of the fluctuation modes ends.  This is the
end of the preheating epoch, and hence we
label the time at which this occurs $\tau_{\rm end}$.  \\
\indent  In order to evaluate $\Sigma (\tau)$, we must calculate 
$U_p^{\prime 2} (0)$ at $p_{\rm max}$.  Using the 
relations~\cite{AbSteg}:
\beqn
\nonumber \zeta \left( z_1 + z_2 \right) &=& \zeta \left( z_1 \right) + 
\zeta \left( z_2 \right) + \frac{1}{2} \frac{ \wp^{\prime} \left( z_1
\right) - \wp^{\prime} \left( z_2 \right)}{ \wp \left( z_1 \right) - \wp
\left( z_2 \right) }, \\
\wp^{\prime} (\omega^{\prime} ) &=& 0 \>\> , \>\> \wp 
(\omega^{\prime}) = e_3 ,
\eeqn
then equations (\ref{e_i}) and (\ref{Up1}) yield
\beqn
\nonumber  U_p^{\prime 2} (0)_{\vert p_{\rm max}} &\simeq& \frac{ 
\wp^{\prime 2} (b_{\rm max})}{ \left[ \wp (b_{\rm max}) - e_3 \right]^2}
\\
&=& \frac{ 2 \left(3 + 9 \tilde{\cal K}^2 \right)^{3/2} - 18 \tilde{\cal
K} \left( 
1 - \tilde{\cal K}^2 \right)}{\left[ 6 - \left(3 - 3 \tilde{\cal K}
\right) \sqrt{3 + 
9 \tilde{\cal K}^2} - 9 \tilde{\cal K} \left(1 - \tilde{\cal K} \right)
\right] } .
\label{Uprime}
\eeqn
Similarly, the quantity
\beq
\omega^2_{\ell_{\rm max}} (0) = \ell_{\rm max}^2 + 3 \psi_o^2 = 
\gamma^2 \left[ 3 + \sqrt{3 + 9\tilde{\cal K}^2} - 3 \tilde{\cal K}
\right] .
\eeq
The backreaction may thus be written:
\beq
\Sigma (\tau) = \frac{3 \gamma^{3/2}}{256 \pi^3} G^{1/2} 
(\tilde{\cal K}) J (\tilde{\cal K}) \sqrt{ \frac{ K (\overline{k})}{q}} 
\frac{1}{\sqrt{\tau}} \exp \left[ \frac{8 \pi \gamma q}{K (\overline{k})}
\tau + O (q^3) \right] , 
\label{Sigma-full}
\eeq
where we have defined
\beqn
\nonumber  G (\tilde{\cal K}) &\equiv& \frac{3 + 9 \tilde{\cal K}^2 - 6 
\tilde{\cal K} \sqrt{ 3 + 9 \tilde{\cal K}^2} }{3 + \sqrt{3 + 9
\tilde{\cal K}^2} - 
3 \tilde{\cal K} } , \\
J (\tilde{\cal K}) &\equiv& \frac{3 + \sqrt{3 + 9 \tilde{\cal K}^2}
\left( 1 + 3 
\tilde{\cal K} + 6 \tilde{\cal K}^2 \right) - 3 \tilde{\cal K} \left( 6 -
3 
\tilde{\cal K} - 2 \tilde{\cal K}^2 \right) }{\left(3 + 9 \tilde{\cal
K}^2 
\right)^{3/2} - 9 \tilde{\cal K} \left( 1 - \tilde{\cal K}^2 \right) } .
\label{GJ}
\eeqn
When $K = 0$, this reduces to the more simple form:
\beq
\Sigma (\tau )_{\vert K=0} = \frac{ \psi_o^{3/2}}{256 \pi^3} \sqrt{ 3 + 
\sqrt{3}} \sqrt{\frac{K (1/\sqrt{2})}{q_{\rm max}}} \frac{1}{\sqrt{\tau}}
\exp 
\left[ \frac{8 \pi \psi_o q_{\rm max}}{K (1/\sqrt{2})} \>\tau \right] .
\label{SigmaK=0}
\eeq
The numerical values $K (1/\sqrt{2}) = 1.854$, $q_{\rm max} = 0.0432$,
and $\lambda = 10^{-12}$ yield for the quantities in equation 
(\ref{back1}),
\beqn
\nonumber N(0) &=& \frac{256 \pi^3}{\sqrt{3 + \sqrt{3}}} \sqrt{ 
\frac{q_{\rm max}}{K (1/\sqrt{2}) }} = 556.995 , \\
\nonumber B(0) &=& \frac{8 \pi q_{\rm max}}{K (1/\sqrt{2})} = 0.586 , \\
\tau_{\rm end}(0) &=& \frac{1}{B(0) \psi_o} \ln  \left( \frac{N(0)}{2
\lambda \sqrt{B (0)}} \right) = \frac{57.266}{\psi_o} .
\label{NBtau}
\eeqn
The quantities $N (\tilde{\cal K})$, $B (\tilde{\cal K})$, and 
$\tau_{\rm end} 
(\tilde{\cal K})$ will all approach these values in the 
limit $\psi_o 
\rightarrow \infty$.  We will calculate the maximum quantitative 
deviation from the flat-space results in section V. \\
\indent  Using the same saddle-point approximation, we may solve for the 
total number of particles produced during preheating.  Within the
resonance band, we will approximate 
\beq
\frac{d \chi_{\ell} }{d\tau} \simeq \left[ \frac{-i \gamma}{2 \omega} F 
(p)\right] \chi_{\ell} (\tau) ,
\eeq
so that near the center of the resonance band (see equation (\ref{Nad})),
\beq
N^{\rm ad}_{\ell} (\tau) \simeq \frac{ \omega_{\ell} (\tau)}{2} \left| 
\chi_{\ell} (\tau) \right|^2 \left[ 1 + \frac{16 \pi^2 q^2 
\gamma^2}{\omega^2 \omega_{\ell}^2 (\tau)} \right] .
\eeq
The total number of particles is
\beq
N^{\rm ad}_{\rm total} (\tau) = \int \frac{d\ell \> \ell^2}{2 \pi^2} \> 
N^{\rm ad}_{\ell} (\tau) .
\label{Ntotal}
\eeq
From equation (\ref{Sigma-res}), using $3 \lambda \Sigma (\tau_{\rm end}) 
\simeq 3 \psi_o^2 / 2$, and approximating $\omega_{\ell} (\tau_{\rm end}) 
\simeq \omega_{\ell} (0)$, this may be written
\beqn
\nonumber N^{\rm ad}_{\rm total} (\tau_{\rm end}) &\simeq& \frac{ 
\gamma^3 \left(1 - 3\tilde{\cal K} \right)}{4 \lambda} \left[ 3 + \sqrt{3
+ 9 
\tilde{\cal K}^2} - 3 \tilde{\cal K} \right]^{1/2} \\
&\times& \left[ 1 + \frac{16 \pi^2 q^2}{K^2 (\overline{k}) \left( 3 +
\sqrt{3 + 9 \tilde{\cal K}^2} - 3 \tilde{\cal K} \right)} \right] .
\label{Ntot2}
\eeqn
Equation (\ref{Ntot2}) confirms that at the end of the preheating epoch,
the 
number of particles produced is nonperturbatively large, $N_{\rm total} 
\propto \lambda^{-1}$.  Thus, for a massless inflaton decaying strictly
into massless inflaton quanta, resonant preheating in an expanding FRW
metric closely resembles the situation for Minkowski spacetime. \\

\section{Comparison of Open and Flat Cases}
\indent  In this section, we will demonstrate that two key approximations 
made above are consistent for the case of preheating with a massless
inflaton, 
and then compare the numerical results for the $K = -1$ and $K = 0$
cases.  
First consider the approximation made in section III that ${\cal H} \ll
\psi$ 
during preheating.  Taking ${\cal H} (0) = 0$, we may write the average
of this 
Hubble parameter over the duration of preheating.  For $K = 0$,
\beq
\overline{\cal H} = \frac{1}{2} {\cal H} (\tau_{\rm end}) = \frac{1}{2 
\tau_{\rm end}} \simeq 10^{-2} \psi_o ,
\eeq
where the last step comes from using equation (\ref{NBtau}).  As
demonstrated below, for the $K = -1$ case, $\tau_{\rm end}$ remains greater
than or equal to its numerical value in the $K = 0$ case, so that
\beq
\overline{\cal H} = \frac{1}{2} {\rm ctnh} (\tau_{\rm end}) < \psi_o 
\eeq
for all $\psi_o^2 > - 2 {\cal K}$, as long as ${\cal C}^2 \leq O (10)$
after renormalization.  Over the 
course of preheating, then, it remains a good approximation to neglect
the 
${\cal H}^2 \psi^2$ term relative to the $\psi^4$ term in the energy
density of 
the zero-mode, equation (\ref{rho2}); when this is done, the energy
density remains proportional to $a^{-4}$ when $m = 0$. \\
\indent  The other main assumption to check concerns the expansion of the 
fluctuating field $\chi$ in subcurvature modes only, when $K = -1$.  For
the 
decay process to be resonant, we require that the period of the Floquet 
solutions remain less than the Hubble time.  In terms of the time-like 
variable $u$, this period is $2 \omega = 2 K (\overline{k})$; in terms of 
conformal time $\eta$, then, this requirement becomes:
\beq
a_o \frac{2 K (\overline{k})}{\gamma {\cal C}} < H_o^{-1} .
\eeq
When $K = -1$, $a_o = \sqrt{2} H_o^{-1}$, so this requirement becomes
\beq
\gamma {\cal C} > 2 \sqrt{2} K(\overline{k}) .
\label{gammaC}
\eeq
The modulus $\overline{k}$ is related to $\gamma$ as (see equation 
(\ref{psisol}))
\beq
\overline{k}^2 = \frac{\gamma^2 {\cal C}^2 - K}{2 \gamma^2 {\cal C}^2}.
\eeq
Equation (\ref{gammaC}) is therefore satisfied for $\gamma {\cal C} \geq
5.3$.
From equations (\ref{chieom3}) and (\ref{p2}), and using $k = {\cal C}
\ell$, this means that 
\beq
k_{\rm min}^2 = \frac{3}{2} \sqrt{\gamma^4 {\cal C}^4 + \frac{K^2 {\cal 
C}^2}{3}} + 2 K \geq \frac{3}{2} \> (5.3)^2 - 2 = 40.1 \>\>,\>\> 
{\rm or} \>\> k_{\rm min} \geq 6.3 .
\eeq
Thus, over the entire range of allowable initial conditions for $\psi_o$, 
preheating with $K = -1$ will only populate subcurvature modes, with 
$k_{\rm min} > 1$. \\
\indent  With this restriction on the combination $\gamma {\cal C}$, we 
may further calculate the quantitative 
deviation for $K = -1$ in the growth of backreaction 
and total number of particles produced from the $K = 0$ case.  Numerical
differences from the $K = 0$ case arise both from the dependence of
$\overline{k}$ (and hence of $K (\overline{k})$ and $q$) on ${\cal K}$, as
well as from the explicit factors of $\tilde{\cal K} = K / (3 \gamma^2
{\cal C}^2)$ in such quantities as $N$, $B$, $\tau_{\rm end}$, and $N^{\rm
ad}_{\rm total}$.  At the minimum allowable value of 
$\gamma {\cal C} = 5.3$,
we have $\overline{k} = 0.72$ (near the flat case of $\overline{k} =
0.707$), or $q = 0.0396$ (near the flat case of $q = 0.0432$).  
With the constants at these values,
\beq
\left| \tilde{\cal K} \right| \leq \frac{1}{3 \gamma^2 {\cal C}^2} = 1.2 
\times 10^{-2} .
\label{tildeK}
\eeq
From the definition of $\gamma^2$, we also have the relation
\beq
\frac{\gamma}{\psi_o} = \left( 1 - 3 \tilde{\cal K} \right)^{-1/2} .
\eeq
This means that the deviation from the $K = 0$ case remains on 
the order of $10^{-2}$.  For example, 
comparing equations (\ref{GJ}) and (\ref{NBtau}),
\beqn
\nonumber G_{\rm max} (\tilde{\cal K}) &=& G (0) + 2.1 \times 10^{-2} , 
\\
\nonumber J_{\rm max} (\tilde{\cal K}) &=& J (0) + 1.0 \times 10^{-2} ,
\\
\nonumber \frac{N_{\rm max} (\tilde{\cal K})}{N (0)} &=& 0.93 , \\
\frac{B_{\rm max} (\tilde{\cal K})}{B (0)} &=& 0.89 .
\label{compare1}
\eeqn
With these values, we see that the duration of preheating also remains
nearly the same in the $K = -1$ and $K = 0$ cases:
\beq
\frac{ \tau_{\rm end} (\tilde{\cal K})}{\tau_{\rm end} (0)} = 1.14 .
\eeq
Finally, the total number of particles produced in the two cases obeys
\beq
\frac{N^{\rm ad}_{\rm total} (\tilde{\cal K})}{N^{\rm ad}_{\rm total}
(0)} = 0.98 .
\eeq
For $\gamma {\cal C} > (\gamma {\cal C})_{\rm min} = 5.3$, $\overline{k}
\rightarrow 1/\sqrt{2}$ and $\tilde{\cal K} \rightarrow 0$, so all
numerical quantities further approach their flat-space values. \\
\indent  Thus, the requirement that the period of the Floquet solutions, 
$2\omega$, 
remain less than one Hubble time greatly restricts the numerical
deviations 
between quantities in the $K = -1$ and $K = 0$ cases, when we consider 
chaotic inflation initial conditions.  For all dynamically-consistent
initial values of $\psi_o$ 
in the open inflation case, the total number of particles
produced during preheating remains within 2 percent of the flat-space
results. \\

\section{Conclusions}
\indent  For a massless inflaton decaying resonantly into massless inflaton
quanta, preheating in expanding FRW spacetimes closely resembles the
Minkowski spacetime case.  With chaotic inflation initial conditions,
furthermore, quantitative differences in the spectra of produced particles
remain small when comparing spatially-flat with open models of inflation. 
In both cases, a single narrow resonance band for comoving wavenumber $k$
exists.  Preheating ends at $\tau_{\rm end}$, when the backreaction
$\lambda \Sigma (\tau)$ due to produced quanta damps the oscillations of
the zero-mode.  Because the Hartree factorization employed here neglects
interactions amongst the final-state bosons, such as re-scatterings between
each other and with the zero-mode, the quantitative values for the spectra 
derived here likely represent an overestimate of their true magnitudes,
though with $\lambda \sim 10^{-12}$, such re-scatterings 
should remain subdominant
effects.  Furthermore, analytic study of the dynamics is useful
for comparing preheating in the $K = 0$ and $K = -1$ cases. \\
\indent  The analytic study also reveals clearly the limits for preheating
with a massive inflaton, when the expansion of the universe is taken into
account.  From the form of the equations of motion in terms of the rescaled
fields and conformal time, it becomes easy to understand why numerical
studies of preheating in an expanding flat-FRW spacetime reveal a lack of
resonant decays for massive inflatons and quanta, for many values of their
couplings. \cite{KT2,ProRo}  A non-zero mass for either the inflaton or 
the quanta would
break conformal invariance.  For chaotic inflation initial conditions as
considered here, solutions for the zero-mode of a
massive inflaton would no longer be simply-periodic or elliptic functions.  
Furthermore, if the mass of the zero-mode were zero but that of the decay
products nonzero, then the equation of motion for the fluctuating field
would include the non-periodic term, $a^2 (\eta) m^2$.  Either of these
situations would mean that the \lq\lq potential" for the fluctuating field 
would no
longer be periodic.  Yet the existence of Floquet solutions, with one or 
more
bands of exponential instabilities, depends upon a periodic potential in
the equation of motion of the fluctuating field.  Only in the limit $g
\varphi_o^2 \gg a^2 (\eta) m^2$, where $m$ is the mass of the produced
quanta and $g$ is the coupling strength between the inflaton and
fluctuating field, will the potential approximate a periodic form.  For the
case of inflaton decay into inflaton bosons, as studied here, limits both
on $\lambda$ and on $m_{\varphi}$ from observed cosmic microwave background
anisotropies make it impossible to satisfy this limit, and {\it no}
preheating may occur when $m_{\varphi} \neq 0$ for either $K = 0$ or $K =
-1$, at least for chaotic inflation initial conditions. \\
\indent  Finally, as demonstrated here, the $K = 0$ and $K = -1$ cases
agree in the limit $\psi_o \gg \left| {\cal K} \right|$, which is not
surprising considering the equations of motion in equations (\ref{psieom})
and (\ref{chieom3}).  For this reason, it would be interesting to compare
the preheating scenario for a symmetry-breaking potential with new
inflation initial conditions, $\psi_o \rightarrow 0$, for both the $K = 0$
and $K = -1$ cases.  The question of a non-thermal restoration of symmetry
with such a potential and initial conditions has been raised for preheating
in both Minkowski and flat-FRW spacetimes~\cite{KLS2} (though see also
\cite{DB205,DBfrw}).  Because the preheating dynamics for such a
model in an expanding open universe differ from those in both of these
spatially-flat cases, such a scenario warrants attention.  This is the
subject of further research. \\


\begin{references}

\bibitem{KLS1} L. Kofman, A. Linde, and A. A. Starobinsky, {\it
	Phys.  Rev. Lett.} {\bf 73}, 3195 (1994).  
\bibitem{STB} Y. Shtanov, J. Traschen, and R. Brandenberger, {\it Phys.
	Rev. D} {\bf 51}, 5438 (1995).  
\bibitem{oldreh} A. Dolgov and A. Linde, {\it Phys. Lett. B}
	{\bf 116}, 329 (1982); L. Abbott, E. Farhi, and M. Wise, {\it
	Phys. Lett. B} {\bf 117}, 29 (1982).  
\bibitem{DB1} D. Boyanovsky, H. J. de Vega,
	R. Holman, D.-S. Lee, and A. Singh, {\it Phys. Rev. D} {\bf 51},
	4419 (1995);  D. Boyanovsky, M.  D'Attanasio, H. J. de Vega, R.
	Holman, D.-S. Lee, and A. Singh, {\it Phys. Rev. D} {\bf 52},
	6805 (1995). 
\bibitem{Yosh} M. Yoshimura, {\it Prog. Theo. Phys.} {\bf 94}, 873
	(1995);  H. Fujisaki, K. Kumekawa, M. Yamaguchi, and M.
	Yoshimura, {\it Phys. Rev. D} {\bf 53}, 6805 (1996);  M. Hotta,
	I. Joichi, S. Matsumoto,
	and M. Yoshimura, {\it Phys. Rev. D} {\bf 55}, 4614 (1997).
\bibitem{DKprd} D. Kaiser, {\it Phys. Rev. D} {\bf 53}, 1776 (1996). 
\bibitem{KT2} S. Khlebnikov and I. Tkachev, {\it Phys. Rev. Lett.} {\bf
	77}, 219 (1996); {\it idem.}, {\it Phys. Lett. B} {\bf 390}, 80
	(1997); 
	{\it idem.}, \lq\lq Resonant decay of Bose condensates," Preprint
	hep-ph/9610477.  
\bibitem{Son} D. Son, {\it Phys. Rev. D} {\bf 54}, 3745 (1996). 
\bibitem{DB205} D. Boyanovsky, H. J. de Vega, R. Holman, and J. F. J.
	Salgado, {\it Phys. Rev. D} {\bf 54}, 7570 (1996);  {\it idem.},
	\lq\lq Preheating and Reheating in Inflationary Cosmology:  A 
	Pedagogical Survey,"  Preprint astro-ph/9609007.   
\bibitem{DBfrw} D. Boyanovsky, D. Cormier, H. J.
	de Vega, R. Holman, A. Singh, and M. Srednicki, \lq\lq Preheating
	in FRW Universes," Preprint hep-ph/9609527; {\it idem.}, \lq\lq
	Scalar Field Dynamics in Friedmann Robertson Walker Spacetimes,"
	Preprint hep-ph/9703327.
\bibitem{AllCamp} R. Allahverdi and B. Campbell, {\it Phys. Lett. B} {\bf
	395}, 169 (1997).
\bibitem{Kim} J. Kim and S. Kim, \lq\lq Quantum Fluctuations and
	Particle Production of Coherently Oscillating Inflaton," Preprint
	hep-ph/9611376.  
\bibitem{Zim} W. Zimdahl, D. Pavon, and R. Maartens, {\it Phys. Rev. D}
	{\bf 55}, 4681 (1997).
\bibitem{kofman} L. A. Kofman, \lq\lq The Origin
	of Matter in the Universe: Reheating after Inflation," Preprint
	astro-ph/9605155.  
\bibitem{KLS2} L. Kofman, A. Linde, and A.
	Starobinsky, {\it Phys.  Rev. Lett.} {\bf 76}, 1011 (1996);  I.
	Tkachev, {\it Phys. Lett. B} {\bf 376}, 35 (1996);  A. Riotto and
	I. Tkachev, {\it Phys. Lett. B} {\bf 385}, 57 (1996); E. Kolb and
	A. Riotto, {\it Phys. Rev. D} {\bf 55}, 3313 (1997). 
\bibitem{KLR} E. Kolb, A. Linde, and A. Riotto, {\it Phys. Rev. Lett} 
	{\bf 77}, 4290 (1996); G. Anderson, A. Linde, and A. Riotto, {\it
	Phys. Rev. Lett} {\bf 77}, 3716 (1996); G. Dvali and A. Riotto,
	{\it Phys. Lett. B}
	{\bf 388}, 247 (1996); M. Yoshimura, \lq\lq Baryogenesis and
	thermal history after inflation," Preprint hep-ph/9605246.
\bibitem{KT1} S. Khlebnikov and I. Tkachev, \lq\lq
	Relic gravitational waves produced after preheating," Preprint
	hep-ph/9701423.  
\bibitem{ProRo} T. Prokopec and T. Roos, {\it Phys. Rev. D} {\bf 55},
	3768 (1997).
\bibitem{DB396} D. Boyanovsky, D. Cormier, H. J. de Vega, and R. Holman,
	{\it Phys. Rev. D} {\bf 55}, 3373 (1997).
\bibitem{Coop} F. Cooper, S. Habib, Y. Kluger, and E. Mottola, \lq\lq
	Nonequilibrium Dynamics of Symmetry Breaking in $\lambda \Phi^4$
	Field Theory," Preprint hep-ph/9610345. 
\bibitem{open} J. Gott, {\it Nature} {\bf 295}, 304
	(1982);  J. Gott and T. Statler, {\it Phys. Lett. B} {\bf 136},
	157 (1984);  M. Bucher, A.  Goldhaber, and N. Turok, {\it Phys.
	Rev. D} {\bf 52}, 3314 (1995);  A.  Linde, {\it Phys. Lett. B}
	{\bf 351}, 99 (1995); 
	A. Linde and A.  Mezhlumian, {\it Phys. Rev. D} {\bf 52}, 6789
	(1995); K. Yamamoto, M.  Sasaki, and T. Tanaka, {\it Astrophys.
	J.} {\bf 455}, 412 (1995).  
\bibitem{Hartree} D. Boyanovsky, D.-S. Lee, and A. Singh,
	{\it Phys.  Rev. D} {\bf 48}, 800 (1993);  D. Boyanovsky, H. J.
	de Vega, and R. Holman, {\it Phys. Rev. D} {\bf 49}, 2769 (1994).
\bibitem{LythWoc} D. H. Lyth and A. Woszczyna, {\it Phys. Rev. D} {\bf
	52}, 3338 (1995).  
\bibitem{GarBel} J. Garc\'{i}a-Bellido, A. R.
	Liddle, D. H. Lyth, and D. Wands, {\it Phys. Rev. D} {\bf 55},
	4596 (1997).
\bibitem{DKsup} D. Kaiser,
	\lq\lq Supercurvature Modes from Preheating in an Open Universe,"
	Preprint astro-ph/9608025.  
\bibitem{BirDavies} N. D. Birrell and P. C. W.
	Davies, {\it Quantum fields in curved space} (New York:
	Cambridge University Press, 1982).  
\bibitem{eigen} E. Lifshiftz, {\it J.
	Phys.} (Moscow) {\bf 10}, 116 (1946); E. Lifshitz and I.
	Khalatinikov, {\it Adv. Phys.} {\bf 12}, 185 (1963);  M. Bander
	and C. Itzykson, {\it Rev. Mod. Phys.} {\bf 38}, 346 (1966);  E.
	Harrison, {\it Rev. Mod. Phys.}
	{\bf 39}, 862 (1967);  L.  Parker and S. Fulling, {\it Phys. Rev.
	D} {\bf 9}, 341 (1974).  
\bibitem{cauchy} M. Sasaki, T. Tanaka, and K.
	Yamamoto, {\it Phys. Rev. D} {\bf 51}, 2979 (1995). 
\bibitem{Bateman} A. Erdelyi {\it et al.}, {\it Higher transcendental
	functions}, volume 2 (New York:  McGraw-Hill, 1953).  
\bibitem{AbSteg} M. Abramowitz and I. Stegun, {\it Handbook of
	Mathematical Functions} (New York:  Dover, 1965). 
\bibitem{Ince} E. L. Ince, {\it Ordinary Differential Equations} (New
	York:  Dover, 1956). 


\end{references}
\end{document}